\documentclass[11pt,twoside]{article}
\usepackage{asp2004}
\usepackage{epsf}
\usepackage{lscape}
\begin{document}
\title{The metallicity dependence of WR wind models}
 \author{G.\ Gr\"afener \& W.-R.\ Hamann} \affil{Department of
  Physics, University of Potsdam, Germany}

\begin{abstract}
  With the advance of stellar atmosphere modelling during the last few
  years, large progress in the understanding of Wolf-Rayet (WR) mass loss
  has been achieved.  In the present paper we review the most
  recent developments, including our own results from hydrodynamic non-LTE
  model atmospheres.  In particular, we address the important question of
  the $Z$-dependence of WR mass loss.  We demonstrate that
  models for radiatively driven winds imply a rather strong dependence on
  $Z$.  Moreover, we point out the key role of the $L/M$-ratio for WR-type
  mass loss.
\end{abstract}
\thispagestyle{plain}

\section{WR-type stellar winds}

WR stars show exceptionally strong winds with mass loss rates of the order
of the single-scattering limit or above. The observed wind efficiency
numbers $\eta = \dot{M}v_\infty/(L_\star/c)$, which denote the ratio
between the wind momentum and the momentum of the radiation field, are
typically in the range of 1-5.  Therefore, {\em if} WR-type winds are
driven by radiation, photons must be used more than once, i.e., after
absorption photons must either be re-distributed into different wavelength
regimes or they must be scattered more than once by overlapping lines in
the extended wind.  Because of their high wind densities, WR\,stars develop
extended atmospheres where the hydrostatic layers are completely hidden to
the observer. In combination with their strong radiation fields,
they develop extreme non-LTE conditions that require sophisticated modeling
techniques.

The key question concerning WR mass loss is {\em why} a star becomes a
Wolf-Rayet star. The observed parameters of early-type O\,stars, namely
their luminosities and effective temperatures, are in principle comparable
with late-type WN stars. O\,supergiants even tend to show a similar He- and
N-enriched surface composition.  Because of their higher mass loss rates,
however, WR stars show dramatically different spectra with strong and broad
wind emissions where O\,stars show tiny photospheric absorption lines.

{\em If} the WR winds are radiation-driven like OB star winds, one would
also expect a similar $Z$-dependence of their mass loss.  In this case,
however, the question must be addressed why WNL stars are so different from
O\,stars.

\section{Monte-Carlo models}

The first wind models which were able to reproduce the high efficiency factors
of WR winds were obtained by Monte-Carlo techniques \citep{luc1:93,spr1:94}.
In this approach the path of single photon energy packets, released at the
wind base, is followed until they escape from the wind.  On their way through
the wind, the photons are scattered by Doppler-shifted spectral lines which
are treated as infinitely narrow pure scattering opacities.  For each
scattering process the energy is determined which is transferred from the
radiation field to the wind material or vice versa.  Finally a 'global'
solution for the mass loss rate is obtained by comparing the total mechanical
wind energy which is gained from the radiation field per time interval with
the overall wind luminosity $\dot{M}v_\infty^2$. In the approach of
\citet{luc1:93} the line opacities are taken from a line list and the atomic
populations and ionization are calculated in a modified nebular approximation.
\citet{spr1:94}, on the other hand, used a purely statistical approach for the
strength and the distribution of the line opacities.

The pilot studies by \citet{luc1:93} and \citet{spr1:94} showed that high wind
momenta with $\eta$ up to 10 can in principle be achieved by multiple line
scattering if enough line opacities are present. Nevertheless, these models do
not explain {\em why} the WR mass loss rates are so high. They rely on an
adopted $\beta$-type velocity structure with a {\em prescribed} terminal wind
velocity $v_\infty$ (note that the wind luminosity goes with $v_\infty^2$). In
particular, the models fail to reproduce the wind acceleration in
the deeper atmospheric layers below a radius of $\sim 2 R_\star$.

Recently, \citet[][see also this volume]{vin1:05} investigated the $Z$-dependence
of WR mass loss by means of similar models as \citet{luc1:93}.  In their
approach, however, the excitation temperatures which enter the calculation of
the ionization structure and the atomic populations, are extracted from a grid
of model atmospheres \citep[see][]{vin1:99}. For the example of a late-type WN
star \citeauthor{vin1:05} find a similar $Z$-dependence as for O-stars with
$\dot{M}\propto Z^{0.86}$ which {\em flattens} at metallicities below
$10^{-4}Z_\odot$ because the line driving is taken over by N and He.  Note
that the mass loss rates at this metallicity are so small ($\sim
10^{-7}M_\odot/\mbox{yr}$) that the star would not be identified as a WR-type
from its spectral appearance. For a WCL model they find a relation with
$Z^{0.66}$ which flattens already at $10^{-3}Z_\odot$.

\begin{figure}[t!]
  \plotfiddle{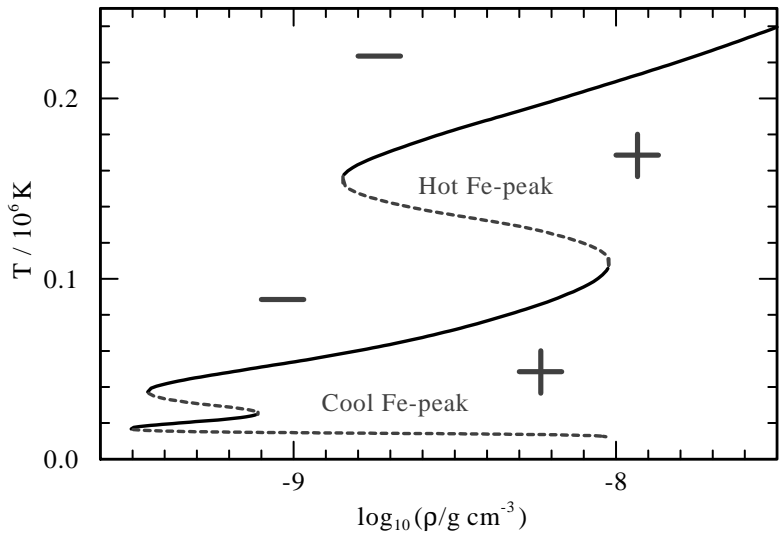}{6.5cm}{0}{75}{75}{-215}{-45}
\caption
{Solution of Eq.\,(\ref{eq:ledd}) in the $\rho$-$T$ plane. The sonic-point
  conditions for an optically thick wind, i.e.\ $\chi_{\rm Ross}=\chi_{\rm
    crit}$ with {\em outward increasing} $\chi_{\rm Ross}$, are fulfilled
  on the solid parts of the curve between 40 and 100\,kK, and above 160\,kK. The
  Rosseland opacities in this plot are taken from the OPAL opacity tables.
  Typical WC star parameters are assumed.
  \label{fig1}
}
\end{figure}

\section{Optically thick wind models}

A approach which is complementary to the previous one is the critical-point
analysis for optically thick winds \citep[e.g.][]{pis1:95,nug1:02}. The
underlying assumption of this method is that, because of their high mass loss
rates, WR atmospheres are so extended that the sonic point of the wind flow is
located at large flux-mean optical depth $\tau_s$. If $\tau_{\rm s}$ is large
enough the radiation transport can be treated in the diffusion limit.  The
expression for the radiative acceleration then simplifies to
\begin{equation}
 \label{eq:arad}
  a_{\rm rad} = \frac{1}{c} \int \chi_\nu F_\nu {\rm d}\nu
  \equiv \chi_{\rm Ross}\frac{L_\star}{4\pi r^2 c}
\end{equation}
where $\chi_{\rm Ross}$ denotes the Rosseland mean opacity which can easily be
obtained e.g.\ from the OPAL opacity tables \citep{igl1:96}. This expression
for $a_{\rm rad}$ is not only easy to evaluate, it also changes the dynamic
wind properties in relation to OB stars. The wind acceleration in a
spherically expanding radiatively driven flow is given by the contributions of
the gravitational attraction, the pressure gradient, and the radiative
acceleration
\begin{equation}
  v\frac{{\rm d}v}{{\rm d}r} = -\frac{MG}{r^2} 
  - \frac{1}{\rho}\frac{{\rm d}p}{{\rm d}r} + a_{\rm rad}.
\end{equation}
When the velocity gradient is extracted from the pressure gradient, this
equation becomes
\begin{equation}
\label{eq:motion}
  \left(v - \frac{a^2}{v}\right) \frac{{\rm d}v}{{\rm d}r} = -\frac{MG}{r^2} 
  + 2\frac{a^2}{r} - \frac{{\rm d}a^2}{{\rm d}r} + a_{\rm rad}
\end{equation}
where $a$ denotes the sonic speed. In the diffusion limit $a_{\rm rad}$ does
not depend on $\frac{{\rm d}v}{{\rm d}r}$. Eq.\,(\ref{eq:motion}) then has a
critical point at the sonic radius $r_{\rm s}$ where $v=a$.  A finite value of
$\frac{{\rm d}v}{{\rm d}r}$ can only be obtained if the right hand side of
Eq.\,(\ref{eq:motion}) is zero at this point.
\begin{equation}
\label{eq:crit}
  0 = -\frac{MG}{r_{\rm s}^2} 
  + 2\frac{a^2}{r_{\rm s}} - \frac{{\rm d}a^2}{{\rm d}r_{\rm s}} + a_{\rm rad}
\end{equation}
In contrast, for the thin winds of OB stars $a_{\rm rad}$ depends on
$\frac{{\rm d}v}{{\rm d}r}$ via Doppler shifts. The critical point of
Eq.\,(\ref{eq:motion}) is then located at a significantly higher speed (the
so-called Abbott speed) which corresponds to the fast radiative-acoustic
wave mode \citep{abb1:80}. For optically thick winds Eq.\,(\ref{eq:crit})
serves as a limiting condition for the mass loss rate. For reasonable wind
parameters the second and third term on the right-hand side of
Eq.\,(\ref{eq:crit}) become negligible so that
\begin{equation}
\label{eq:ledd}
  \frac{MG}{r_{\rm s}^2} \approx a_{\rm rad}(r_{\rm s})
  \equiv \chi_{\rm crit}\frac{L_\star}{4\pi r_{\rm s}^2 c}.
\end{equation}
The Eddington limit, referring to the Rosseland mean opacity, is therefore
crossed at the sonic point. The critical value $\chi_{\rm crit}$ for the
Rosseland mean opacity can be calculated directly from the given $L/M$ ratio
for a specific object.

In Fig.\,\ref{fig1} we plot the the solution of Eq.\,(\ref{eq:ledd}), i.e.\ 
the relation between density and temperature where $\chi_{\rm Ross}(\rho,
T)= \chi_{\rm crit}$, for the example of a typical WC star.  In the
hydrostatic layers below $r_{\rm s}$, the radiative force must be lower
than the Eddington value.  $\chi_{\rm Ross}$ therefore has to increase
outward, with decreasing density. As shown in Fig.\,\ref{fig1} this
condition is fulfilled at the hot edges of the two Fe opacity peaks, at
temperatures in the range of 40--100\,kK and above 160\,kK.  The resulting
mass loss rates on these parts of the curve are given by $\dot{M}= 4\pi
R_\star^2\rho\,a$, where the sonic speed $a$ depends on $T$ and the chemical
composition of the wind material.

To estimate the actual density and temperature at the sonic point, however, a
further constraint is needed. At this point \citet{nug1:02} utilize an
approximate relation between temperature and optical depth by \citet{luc1:71}
\begin{equation}
\label{eq:lucy}
 T^4(r) = \frac{3}{4}T_{\rm eff}^4 \left( \tau'(r)+\frac{4}{3}W(r)\right)
\end{equation}
with the modified optical depth $\tau'$ and the dilution factor $W$ which is
close to unity in our case.  $\tau'$ is obtained from the assumption that the
outer wind is driven by radiation. For this purpose it is necessary to take
the density and velocity structure above the sonic point into account.  Again,
the results rely on the adopted velocity distribution $v(r)$, where a
$\beta$-type velocity law with a terminal wind velocity in the observed range
of 1000--3000\,km/s is used.

The analysis by \citeauthor{nug1:02} reveals important insights how the
mass loss of WR stars might be adjusted.  They find that the observed WR
mass loss rates are in agreement with the optically thick wind assumption,
with two distinct sonic-point temperature regimes.  The cool regime with
typical $T_{\rm s}$ in the range of 40--100\,kK and $\tau_{\rm s} = 3-10$
corresponds to late-type WN stars, and the hot regime with $T_{\rm s}
\approx 160\,{\rm kK}$ and $\tau_{\rm s} = 1-30$ to early-type WN and WC
stars.  For the mass loss of early-type WN stars, Nugis (these proceedings)
finds a $Z$-dependence with very steep exponents in the range 1-1.5,
dependent on the stellar mass. When the mass loss goes down at low
metallicities, the validity of the optically thick wind assumption is
however questionable.

\section{PoWR hydrodynamic model atmospheres}

From Monte-Carlo models we have learned that the acceleration of the outer
part of WR winds, namely their high wind performance numbers, can in principle
be explained by multiple line-scattering. In addition, the critical-point
analysis by \citeauthor{nug1:02} revealed that the mass loss rates are
presumably adjusted by the radiative driving of Fe-peak opacities at large
optical depth. As a consequence we expect two regimes of WR mass loss
corresponding to the two Fe-opacity peaks at cool and hot temperatures.
Moreover, in addition to a $Z$-dependence, we expect a strong dependence of
$\dot{M}$ on the $L/M$ ratio.

Regardless of these important findings, fundamental questions still remain
open. All previous models rely on pre-defined velocity distributions
$v(r)$. It is therefore not clear if the high mass loss rates can be
maintained throughout the {\em whole} wind by radiation pressure alone. In
reality, the terminal wind velocity adjusts by an interplay between the
acceleration in the outer parts of the wind and the limiting condition at the
critical point. Moreover, the optically thick wind assumption, i.e., the
question whether the diffusion limit is valid at the critical point or if the
radiation field is affected by Doppler shifts, remains to be verified.
Finally, spectral analyses and time resolved spectroscopy show direct evidence
for a clumped wind structure, but the effects of clumping are not included in
any previous stellar wind models.

The new Potsdam Wolf-Rayet (PoWR) hydrodynamic model atmospheres
\citep[see][]{gra1:05,ham1:03,koe1:02,gra1:02} address exactly the
questions explained above.  By a combination of line-blanketed non-LTE
model atmospheres with the equations of hydrodynamics, \citet{gra1:05}
obtained the first fully self-consistent models for WR winds.  In these
models $\dot{M}$, $v(r)$, $T(r)$, and the full set of non-LTE population
numbers are computed consistently with the radiation field in the co-moving
frame (CMF), i.e., the radiation transport, the equations of statistical
equilibrium, the energy equation, and the equation of motion are
simultaneously solved.  Through the exact solution of the radiation
transport no simplifying assumptions concerning the radiative acceleration
are necessary.  $a_{\rm rad}$ is obtained from integrating the
product of opacity and flux over frequency (see Eq.\,{\ref{eq:arad}), which
  are both calculated on a fine frequency grid in the co-moving frame of
  reference.  Moreover, clumping is taken into account in the limit of
  optically thin clumps. The resultant models describe the conditions in
  WR\,atmospheres in a realistic way, and provide synthetic spectra, i.e., they
  allow a direct comparison with observational material.  Nevertheless,
  simplifying assumptions are still necessary.  These are especially the
  assumption of a constant Doppler broadening velocity throughout the
  atmosphere, and the omission of opacities, partly due to the neglect of
  trace elements like Ne, Ar, S, or P and partly due to incompleteness of
  the available data.
  
  In the following we present PoWR models for both expected temperature
  regimes. `Hot' models with core temperatures of $T_\star=140\,{\rm kK}$ for
  early-type WC stars on the He-main sequence, and `cool' models for late-type
  WN stars in the range of $T_\star=30$--60\,kK. For the latter we also
  investigate the detailed $Z$-dependence.

\begin{figure}[t!]
\plotfiddle{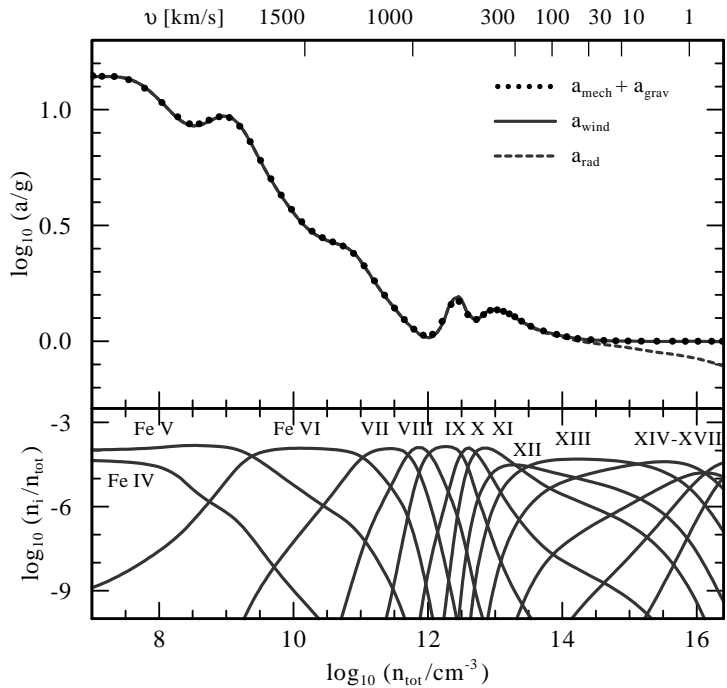}{9cm}{0}{70}{70}{-160}{-50}
\caption{
  Top: Acceleration within our hydrodynamic WC star model in units of
  the local gravity.  The wind acceleration $a_\mathrm{wind}$ due to
  radiation + gas pressure is in precise agreement with the mechanical +
  gravitational acceleration $a_\mathrm{mech} + a_\mathrm{grav}$.  Bottom:
  Fe-ionization structure.
  \label{acc}
}
\end{figure}

\begin{figure}[t!]
\plotfiddle{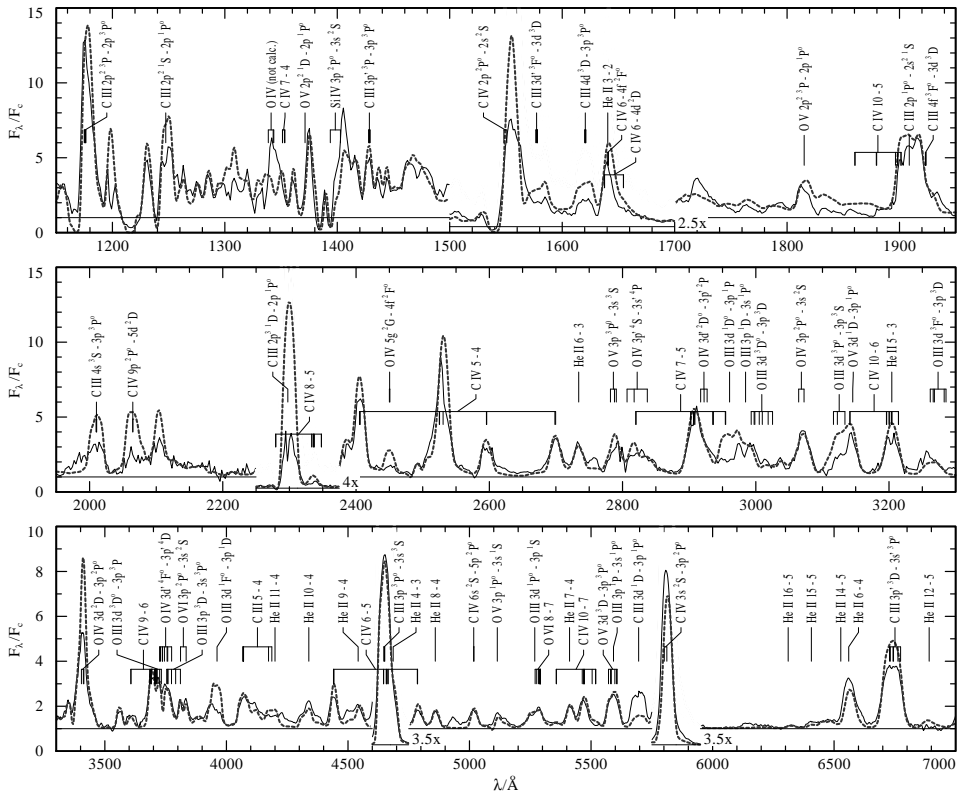}{10.2cm}{0}{46}{46}{-193}{-80}
\caption{
  WR\,111: Comparison of the synthetic spectrum from our hydrodynamic
  $13.6\,M_\odot$ WCE model (dashed line, grey) with observations.
  \label{spec}
}
\end{figure}

\begin{figure}[t!]
\plotfiddle{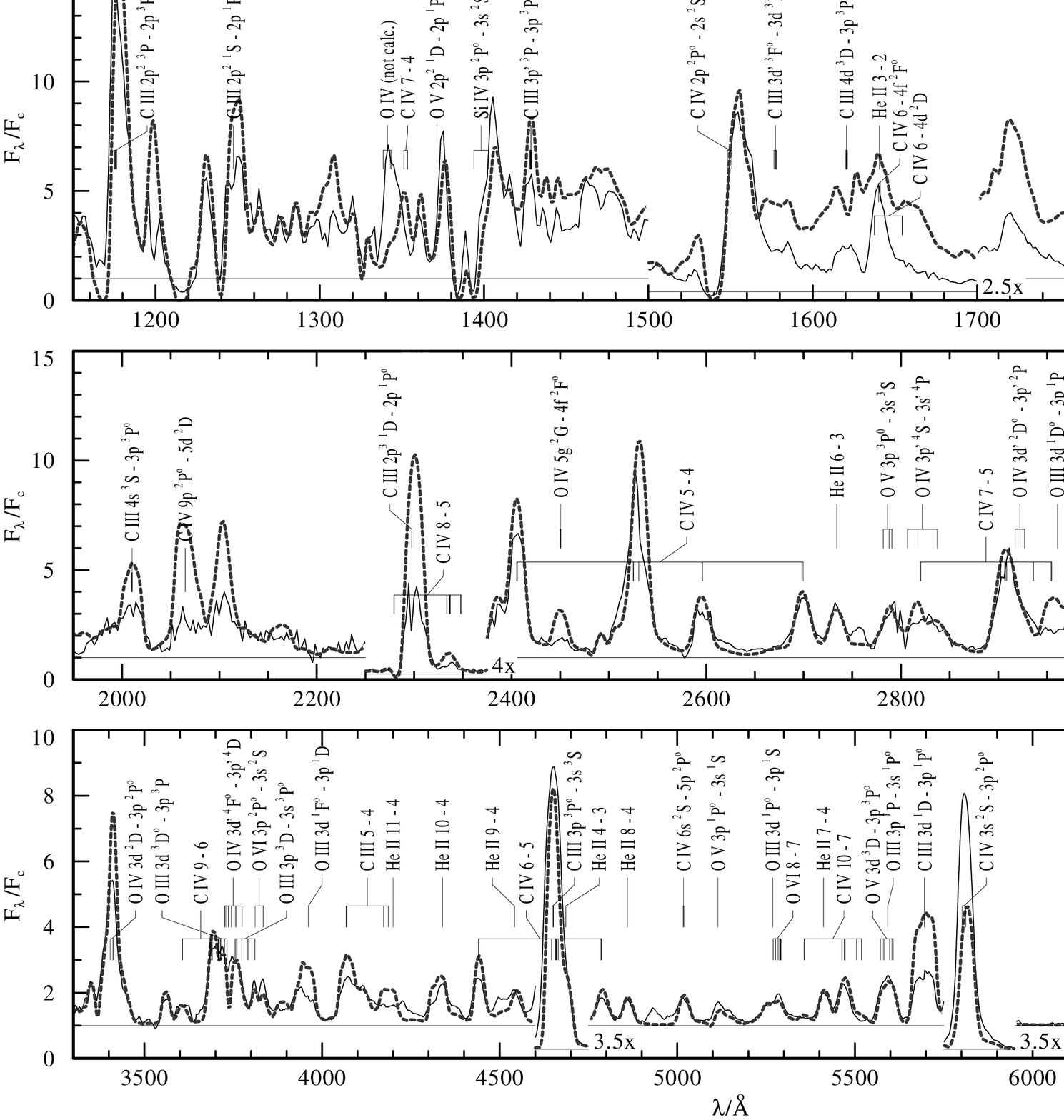}{10.2cm}{0}{46}{46}{-193}{-80}
\caption{
  Same as Fig.\,\ref{spec}, but a reduced mass of $12\,M_\odot$ has been
  adopted in the hydrodynamic calculation.
\label{spec2}
}
\end{figure}

\subsection{Hot objects: early-type WC stars}

WC stars are expected to be chemically homogeneous stars in the phase of
central He-burning. Because of their strong mass loss they show
exceptionally strong emission lines of He, C and O.  They are an ideal
target for our model calculations because they obey a mass-luminosity
relation which has been derived theoretically e.g.\ by \citet{lan1:89}.
This means that the number of free parameters in our calculation is reduced
by one. When the luminosity and the chemical composition for a specific
object are given, e.g.\ from spectral analyses, only the stellar core
radius $R_\star$ or, equivalently, the corresponding effective core
temperature $T_\star$ remains. In fact, this parameter is also given by
spectral analyses.  There are, however, striking discrepancies between the
observed $T_\star$ (50--90\,kK) and the theoretically expected core
temperatures from stellar structure calculations (120--150\,kK).

There are in principle two ways how this discrepancy might be resolved.
First, due to their strong winds, the WC star atmospheres are so extended
({\em above} the sonic point) that the photosphere is located far out in the
wind (at $\sim 3\,R_\star$).  The hydrostatic layers are therefore not
directly observable.  Second, He-burning stars might develop an extended
convective envelope {\em below} the sonic point.  Based on static models,
\citet{ish1:99} have shown that this might be the case for very high
luminosities and/or metallicities. Note, however, that the extension in
these models occurs when the Eddington limit is approached close to the
`hot' Fe-opacity peak. Presumably because of the hydrostatic assumption,
these models form an extended convective zone instead of launching an
optically thick wind. Our hydrodynamic wind models now have the potential
to discriminate between these two possibilities.  On one hand they provide
the sub-photospheric wind structure, on the other hand they can show under
which conditions an optically thick wind actually develops.

The first self-consistent PoWR wind models for an early-type WC star are
described in \citet{gra1:05}. These calculations are based on a previous
spectral analysis of the WC\,5 prototype WR\,111 \citep{gra1:02}, where an
effective core temperature of $T_\star=85\,{\rm kK}$ was obtained.  For the
hydrodynamic calculations, the models now had to be extended by the Fe-peak opacities,
namely the Fe M-shell ions Fe\,{\sc ix--xvii}.  First tests however
demonstrated that {\em no wind-driving is possible for `cool' stellar
  temperatures around 85\,kK}. The models in this temperature regime showed
a strong deficiency in the radiative acceleration directly above the sonic
point. If, on the other hand, $T_\star$ is increased to a value of 140\,kK
(at fixed luminosity) this problem is resolved.

Remaining problems in intermediate wind layers (at velocities around 1000
km/s) are compensated by the choice of a relatively high clumping factor
($D=50$). Wind clumping reduces the {\em observed} mass loss rates by a
factor of $\sqrt{D}$ because it enhances recombination processes. When the
wind density is reduced by this factor, the typical WR emission line
spectrum remains constant. Our models now show that the radiative force
behaves similar, i.e., when clumping is increased and $\dot{M}$ decreased
the {\em force} remains similar which means that $a_{\rm rad}$ {\em
  increases}.  Note that \citet{gra2:03} have shown that O star winds show
the same behavior.

For the hydrodynamic WCE model we assume a stellar mass of
$13.6\,M_\odot$ according to the relation by \citet{lan1:89} for a
luminosity of $L_\star = 10^{5.45}L_\odot$. The resulting wind acceleration
is plotted in Fig.\,\ref{acc} together with the Fe-ionization structure.
With the obtained mass loss rate of $\dot{M}=10^{-5.14}\,M_\odot/{\rm yr}$
and the terminal wind velocity of $v_\infty=2010\,{\rm km}/{\rm s}$ the
observed spectrum of WR\,111 is convincingly reproduced (Fig.\,\ref{spec}).
Nevertheless, the weakness of the electron scattering wings (e.g.\ of
C\,{\sc iv}\,5808) indicates that the choice of $D=50$ is too high.
Although the clumping factors can only be determined very roughly,
WC\,stars usually show values of the order of $D=10$.  Our models might
therefore underestimate the mass loss rate by a factor of $\sqrt{5}$,
presumably due to the incompleteness of opacities.

Concerning the temperature regime where the sonic point is located, our
models are in line with the results of \citet{nug1:02}.  However, because
of the higher effective core temperature in our model, the temperature of
$T_{\rm s} = 199\,{\rm kK}$ is reached at smaller optical depth
$\tau_{\rm s} = 5.4$ (cf.\ Eq.\,\ref{eq:lucy}). Furthermore, our models
show an extreme sensitivity to the $L/M$ ratio. A change of the stellar
mass from 13.6 to $12\,M_\odot$ increases the mass loss by a factor 1.5.
As shown in Fig.\,\ref{spec2}, the resulting model spectrum corresponds
to a later spectral type.

\begin{figure}[t!]
\plotfiddle{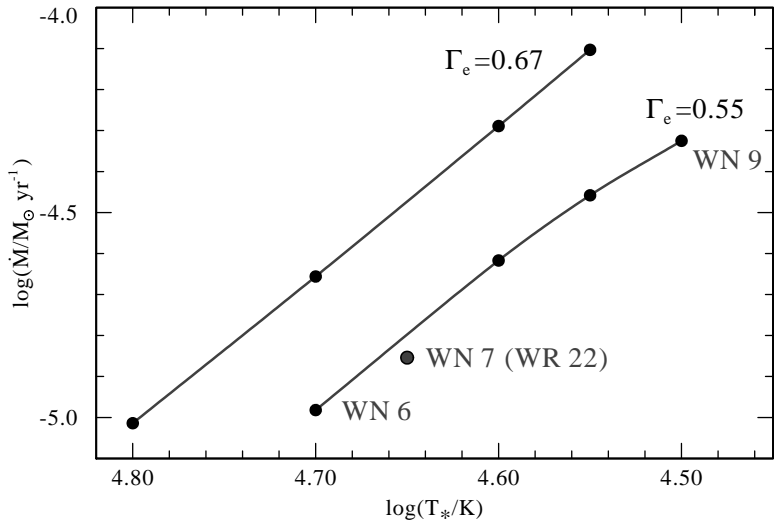}{6.5cm}{0}{50}{50}{-195}{-25}
\caption{
  WNL mass loss rates for different stellar temperatures and masses
  (indicated by the Eddington factor $\Gamma_{\rm e}$). The model for
  WR\,22 is calculated with an enhanced hydrogen abundance (see text).
  \label{mdot-wnl}
}
\end{figure}

\subsection{Cool objects: WNL stars}

Hamann et.\ al (this volume) have analyzed a large part of the galactic WN
stars by means of a grid of line-blanketed PoWR models \citep{ham1:04}. The
stars in this sample divide into two distinct groups. The first group
consists of a mixture of early- to intermediate-type WN stars with
luminosities below $10^6\,L_\odot$. The second group consists of late-type
WN stars (WN\,6--9) with luminosities above $10^6\,L_\odot$. The latter are
located to the right of the main sequence (at $T_\star = 35$--55\,kK) and
show hydrogen at their surface. The extreme brightness of these stars
already implies high $L/M$ ratios.  We have therefore calculated a grid of
WNL models with a fixed luminosity of $10^{6.3}\,L_\odot$ and core
temperatures in the range of 30--60\,kK.  For the stellar masses, values of
67 and $55\,M_\odot$ were adopted, corresponding to Eddington factors
$\Gamma_{\rm e}\equiv\chi_{\rm e}L_\star/4\pi c G M_\star$ of 0.55 and
0.67.  Typical WR surface abundances of $X_{\rm H}=0.2$ and $X_{\rm
  N}=0.015$, and a clumping factor of $D=10$ are assumed.

The resulting mass loss rates are plotted in Fig.\,\ref{mdot-wnl}.  As
expected for optically thick winds, the mass loss shows a strong dependence
on the ratio $L/M$ or equivalently on $\Gamma_{\rm e}$. Moreover, the mass
loss increases with decreasing $T_\star$, which can also be understood in
the framework of the optically thick wind theory. At cooler $T_{\rm eff}$ a
higher optical depths is needed at the sonic point to reach the demanded
temperatures (see Eq.\,\ref{eq:lucy}). Therefore higher wind densities are
needed. The obtained synthetic spectra reflect the observed sequence of WNL
spectral subtypes, starting with WN\,6 at 55\,kK to WN\,9 at 31\,kK.

To verify our assumption of a high $L/M$ ratio for WNL stars we have
performed a more detailed investigation of the WN\,7 component in the
eclipsing WR+O system WR\,22. The O\,star in this system is so faint that
the flux distribution in the optical and the UV is dominated by the WR
star.  Nevertheless, mass estimates are available from orbital solutions
for the marginally visible O\,star absorption features. On this way values
of $M_{\rm WR}\,\sin^3 i = 55\pm 7\,M_\odot$ \citep{sch1:99} and $72\pm
3\,M_\odot$ \citep{rau1:96} have been determined, both with a ratio of
$M_{\rm WR}/M_{\rm O} = 2.7$.

\begin{figure}[t!]
\plotfiddle{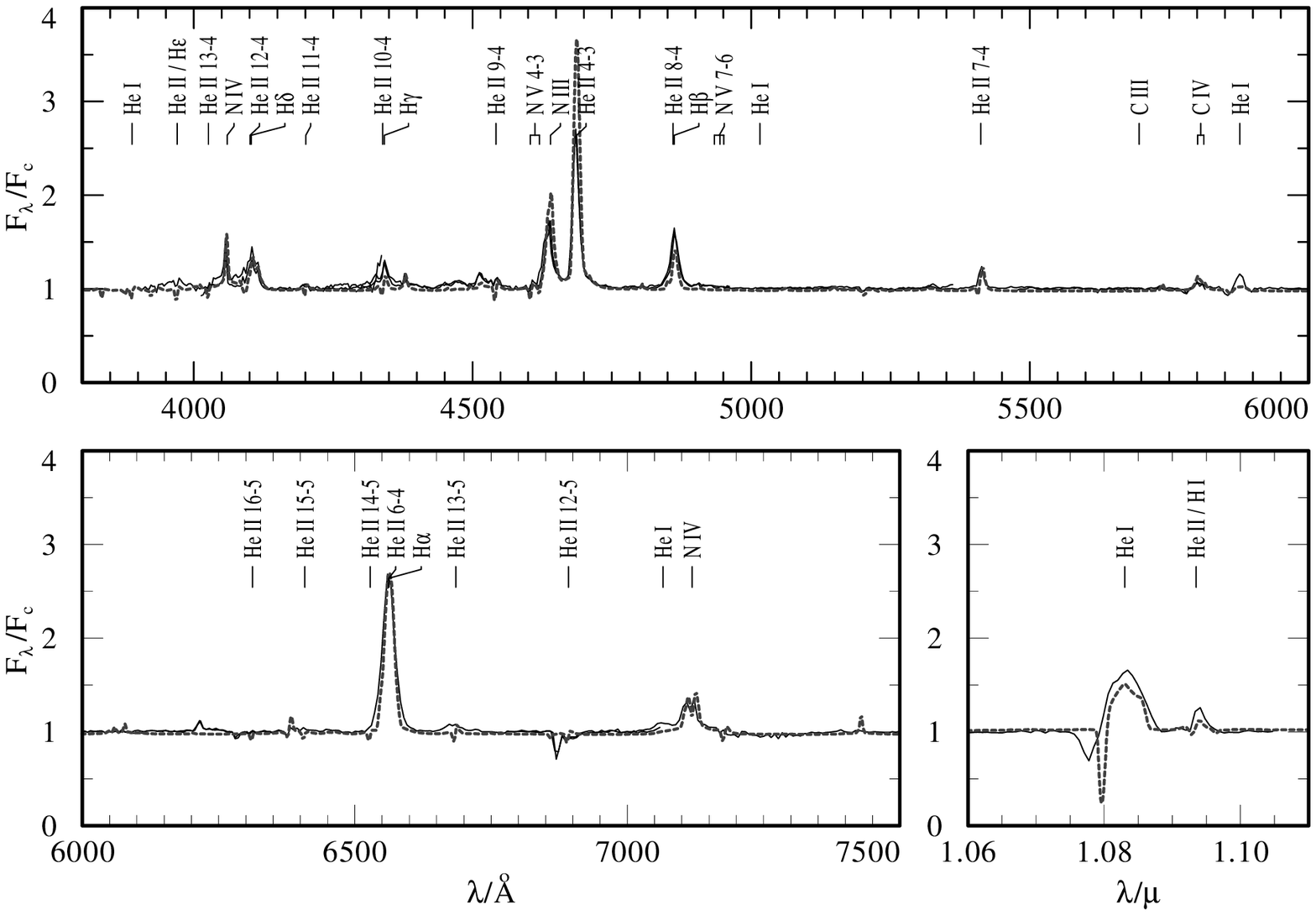}{8.2cm}{0}{65}{65}{-210}{-30}
\caption{
  WR\,22: Comparison of the synthetic spectrum from our hydrodynamic
  model (dashed line, grey) with observations.
  \label{wr022}
}
\end{figure}

With our self-consistent wind models we obtain a reasonable spectral fit
for WR\,22 with $\Gamma_e=0.67$, a slightly enhanced hydrogen abundance
$X_{\rm H} = 0.4$, and a stellar temperature of 45\,kK (see
Fig.\ref{wr022}).  With the standard distance modulus of 12.1\,mag to
Car\,OB1, our model with $L_\star=10^{6.3}\,L_\odot$ exactly reproduces the
observed flux distribution.  The corresponding stellar mass of
$78\,M_\odot$ is in excellent agreement with the mass determination by
\citet{rau1:96}.  Note, however, that the resulting stellar mass depends on
$X_{\rm H}$ (because $\Gamma_e$ depends on $X_{\rm H}$) {\em and} on the exact stellar
luminosity, i.e.\ the adopted distance.  With a distance modulus of
12.55\,mag, as determined by \citet{mas1:93} for the Carina OB clusters
Tr\,14 and Tr\,16, we would obtain an even higher luminosity and therefore
an even higher $L/M$ ratio. In any case, WR\,22 has a luminosity equal or
above $10^{6.3}\,L_\odot$ and is located rather close to the Eddington
limit.

\begin{figure}[t!]
\plotfiddle{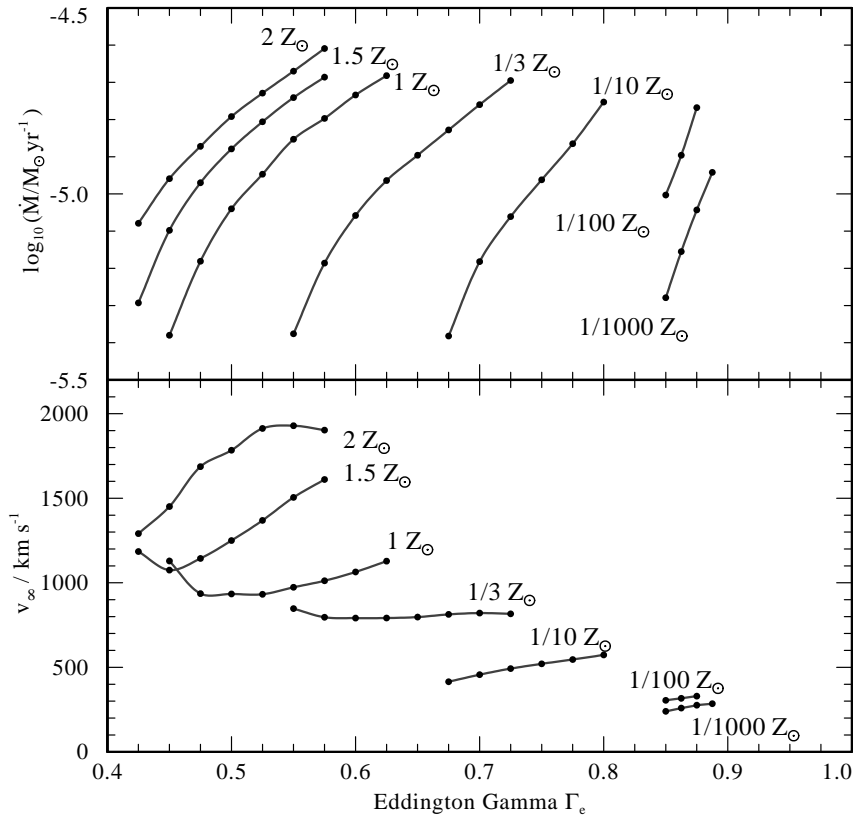}{10.5cm}{0}{45}{45}{-190}{-30}
\caption{
  $Z$-dependence for WNL stars: Mass loss rates (top) and terminal wind
  velocities (bottom) as obtained from our grid of hydrodynamic models.
\label{wnl-z}
}
\end{figure}

\subsection{WNL stars: metallicity-dependence}

The WR-type winds in our models are predominantly driven by
radiation pressure on iron line opacities.  We therefore expect a strong
dependence of WR mass loss on $Z$. To investigate this dependence we have
prepared a model grid for late-type WN stars which is based on our model
for WR\,22.  Because we already know about the importance of the $L/M$
ratio, we have varied the Eddington factor $\Gamma_e$ (or equivalently the
stellar mass) in addition to the metallicity $Z$ (note that we scale all
metals {\em including nitrogen} with $Z$).  As shown in Fig.\ref{wnl-z},
our models indeed show a strong $Z$-dependence.  For models with {\em fixed
 L/M ratio} we find rather similar exponents as \citet{vin1:05}
($\dot{M} \propto Z^{0.86}$).  In addition, however, our models show a
strong dependence on $\Gamma_e$. In particular, we find that optically
thick winds with high WR-type mass loss rates {\em can even be maintained
  at extremely low Z}, if the star manages to come close enough to the
Eddington limit.

The obtained terminal wind velocities are shown in Fig.\ref{wnl-z}
(bottom).  First, our models clearly predict decreasing wind velocities
with decreasing $Z$.  This finding is generally confirmed by observations
\citep[see e.g.][]{con1:89,cro2:00}, although the observational evidence is not as
clear as our prediction. Moreover, our models show a rather striking
behavior because the terminal velocities stay constant, or even increase
with increasing mass loss rate.  Obviously, our dense wind models violate
the well-established wind momentum-luminosity relation for OB star winds
\citep[e.g.][]{kud1:99}, which would imply $\dot{M}v_\infty = const$.
Only at the lowest densities the terminal wind velocity starts to increase
as expected. A closer inspection of the models shows that the changes at
high wind densities are related to changes in the ionization structure. It
therefore seems that the velocity structure is dictated by ionization
effects rather than by CAK-type wind physics. This is in line with our
result that, in contrast to standard assumptions, the {\em effective} force
multiplier parameter $\alpha$ shows values close to zero in our WC star
model \citep[see][]{gra1:05}. Also the present WNL models show a tendency
towards low $\alpha$-values ($\approx 0.2$), although not as severe as our
WC model.

\section{Conclusions}

We conclude that models for radiatively driven Wolf-Rayet winds imply a
strong $Z$-dependence of the mass loss. In addition, the $L/M$-ratio turns
out to be at least equally important. Proximity to the Eddington-limit, or
over-luminosity, might even be pre-requisite for WR-type mass loss.  This
would naturally explain why He-burning stars {\em and} extremely luminous
stars tend to show WR spectra. Moreover it implies that rotation (which is
not yet included in our models) might play an important role. First,
rotation enhances the $L/M$ ratio in the H-burning phase by mixing hydrogen
into the stellar core. Second, rotation reduces the effective gravity at
the stellar surface and thereby also enhances the effective Eddington
factor. Most importantly, we can state that stellar winds from WR\,stars
can now be explained by self-consistent hydrodynamic models.


\begin{thebibliography}
\expandafter\ifx\csname natexlab\endcsname\relax\def\natexlab#1{#1}\fi

\bibitem[{{Abbott}(1980)}]{abb1:80}
{Abbott}, D.~C. 1980, ApJ, 242, 1183

\bibitem[{{Conti} {et~al.}(1989){Conti}, {Garmany}, \& {Massey}}]{con1:89}
{Conti}, P.~S., {Garmany}, C.~D., \& {Massey}, P. 1989, ApJ, 341, 113

\bibitem[{{Crowther}(2000)}]{cro2:00}
{Crowther}, P.~A. 2000, A\&A, 356, 191

\bibitem[{{Gr{\" a}fener} \& {Hamann}(2003)}]{gra2:03}
{Gr{\" a}fener}, G. \& {Hamann}, W.-R. 2003, in IAU Symp., Vol. 212, A Massive
  Star Odyssey: From Main Sequence to Supernova, ed. K.~A. {van der Hucht},
  A.~{Herrero}, \& E.~{C\'esar} (San Francisco: ASP), 190

\bibitem[{{Gr{\" a}fener} \& {Hamann}(2005)}]{gra1:05}
{Gr{\" a}fener}, G. \& {Hamann}, W.-R. 2005, A\&A, 432, 633

\bibitem[{{Gr\"afener} {et~al.}(2002){Gr\"afener}, {Koesterke}, \&
  {Hamann}}]{gra1:02}
{Gr\"afener}, G., {Koesterke}, L., \& {Hamann}, W.-R. 2002, A\&A, 387, 244

\bibitem[{{Hamann} \& {Gr{\"a}fener}(2003)}]{ham1:03}
{Hamann}, W.-R. \& {Gr{\"a}fener}, G. 2003, A\&A, 410, 993

\bibitem[{{Hamann} \& {Gr{\"a}fener}(2004)}]{ham1:04}
{Hamann}, W.-R. \& {Gr{\"a}fener}, G. 2004, A\&A, 427, 697

\bibitem[{{Iglesias} \& {Rogers}(1996)}]{igl1:96}
{Iglesias}, C.~A. \& {Rogers}, F.~J. 1996, ApJ, 464, 943

\bibitem[{{Ishii} {et~al.}(1999){Ishii}, {Ueno}, \& {Kato}}]{ish1:99}
{Ishii}, M., {Ueno}, M., \& {Kato}, M. 1999, PASJ, 51, 417

\bibitem[{{Koesterke} {et~al.}(2002){Koesterke}, {Hamann}, \& {Gr{\"
  a}fener}}]{koe1:02}
{Koesterke}, L., {Hamann}, W.-R., \& {Gr{\" a}fener}, G. 2002, A\&A, 384, 562

\bibitem[{{Kudritzki} {et~al.}(1999){Kudritzki}, {Puls}, {Lennon}, {Venn},
  {Reetz}, {Najarro}, {McCarthy}, \& {Herrero}}]{kud1:99}
{Kudritzki}, R.~P., {Puls}, J., {Lennon}, D.~J., {et~al.} 1999, A\&A, 350, 970

\bibitem[{{Langer}(1989)}]{lan1:89}
{Langer}, N. 1989, A\&A, 210, 93

\bibitem[{{Lucy}(1971)}]{luc1:71}
{Lucy}, L.~B. 1971, ApJ, 163, 95

\bibitem[{{Lucy} \& {Abbott}(1993)}]{luc1:93}
{Lucy}, L.~B. \& {Abbott}, D.~C. 1993, ApJ, 405, 738

\bibitem[{{Massey} \& {Johnson}(1993)}]{mas1:93}
{Massey}, P. \& {Johnson}, J. 1993, AJ, 105, 980

\bibitem[{{Nugis} \& {Lamers}(2002)}]{nug1:02}
{Nugis}, T. \& {Lamers}, H.~J.~G.~L.~M. 2002, A\&A, 389, 162

\bibitem[{{Pistinner} \& {Eichler}(1995)}]{pis1:95}
{Pistinner}, S. \& {Eichler}, D. 1995, ApJ, 454, 404

\bibitem[{{Rauw} {et~al.}(1996){Rauw}, {Vreux}, {Gosset}, {Hutsemekers},
  {Magain}, \& {Rochowicz}}]{rau1:96}
{Rauw}, G., {Vreux}, J.-M., {Gosset}, E., {et~al.} 1996, \aap, 306, 771

\bibitem[{{Schweickhardt} {et~al.}(1999){Schweickhardt}, {Schmutz}, {Stahl},
  {Szeifert}, \& {Wolf}}]{sch1:99}
{Schweickhardt}, J., {Schmutz}, W., {Stahl}, O., {Szeifert}, T., \& {Wolf}, B.
  1999, A\&A, 347, 127

\bibitem[{{Springmann}(1994)}]{spr1:94}
{Springmann}, U. 1994, A\&A, 289, 505

\bibitem[{{Vink} \& {de Koter}(2005)}]{vin1:05}
{Vink}, J.~S. \& {de Koter}, A. 2005, A\&A, 442, 587

\bibitem[{{Vink} {et~al.}(1999){Vink}, {de Koter}, \& {Lamers}}]{vin1:99}
{Vink}, J.~S., {de Koter}, A., \& {Lamers}, H.~J.~G.~L.~M. 1999, A\&A, 350, 181

\end{thebibliography}

\end{document}